\documentclass[aps,prb,twocolumn,groupedaddress]{revtex4}
\usepackage{graphicx}

\begin{document}

\newcommand{\be}{\begin{equation}}
\newcommand{\ee}{\end{equation}}
\newcommand{\bea}{\begin{eqnarray}}
\newcommand{\eea}{\end{eqnarray}}
\newcommand{\nt}{\narrowtext}
\newcommand{\wt}{\widetext}

\title{Long-range-correlated disorder in graphene}

\author{D. V. Khveshchenko}

\address{Department of Physics and Astronomy, University of North
Carolina, Chapel Hill, NC 27599}

\begin{abstract}
We study transport of two-dimensional quasi-relativistic 
electronic excitations in graphene in the presence of 
static long-range-correlated random scalar and vector potentials. 
Using a combination of perturbation theory and path-integral techniques, we estimate scattering rates which control Drude conductivity, magneto-transport, and Friedel oscillations in the ballistic regime of large quasiparticle energies. We also discuss properties of zero-energy states and pertinent localization scenarios.
\end{abstract}
\maketitle

\nt
In recent years, the theory of electron localization in two dimensions
(2D) has been extended to the situations where, instead
of (or in addition to) a random potential (RP), there exists a random
magnetic field (RMF). This type of problems emerge in the context of compressible Quantum Hall states \cite{hlr}, finite-temperature dynamics of spin liquid states in Mott insulators \cite{RMF}, and vortex line liquid phase in high-$T_c$ cuprates \cite{franz}, to name a few. 

For example, in the latter case fermionic excitations 
possess quasi-relativistic (albeit not necessarily 
rotationally invariant) 
kinematics, and the roles of RP and RMF are played by the Doppler shift
due to a circulating supercurrent and concomitant Berry phase, respectively \cite{franz}.

Another relevant application of this disorder model can be found
in graphene which has received a lot of attention lately.
Although a number of different (anti)localization-related phenomena have already been discussed \cite{ando,aleiner}, these studies were largely limited to the case of short-ranged (albeit, possibly, arbitrarily strong) disorder.

Recently, the effect of long-range Coulomb impurities residing in the $SiO_2$ substrate was invoked \cite{nomura} to explain the experimentally observed linear dependence of the conductivity on electron density \cite{exp}. 

It has also been pointed out that in graphene, besides the Coulomb ("scalar") RP, there exists a "vector" (RMF) disorder 
representing the effect of disclinations (isolated pentagon- and heptagon-rings),  dislocations (pairs of adjacent pentagons and heptagons), and Stone-Wales defects (double pairs) \cite{guinea}. The presence of such structural defects in free-standing graphene
is inevitable due to the intrinsic thermodynamic instability of 
2D crystals.

Nevertheless, in most of the previous work this RMF was repeatedly treated as being short-range in terms of the vector potential (rather than the magnetic field itself),
which simplification facilitates the use of a powerful machinery of 
renormalization group \cite{guinea} and 2D conformal field theory \cite{Ludwig}.

In the present Letter, we compare the effects of long-range-correlated RP and RMF on electron transport in graphene and contrast the results with those pertaining to the conventional 2DEG with parabolic electron dispersion.

The Dirac-like quasiparticles near two conical points ($K$ and $K^\prime$) in the hexagonal Brillouin zone of graphene cn be described in terms of the (retarded) Green function
\be 
{\hat G}^R(\omega,{\bf p})={(\epsilon+\Sigma^R){\hat 
\gamma}_0+v{\bf {\hat \gamma}}{\bf p}
\over {(\epsilon+\Sigma^R)^2-v^2{\bf p}^2}} 
\ee 
where $v$ is the Fermi velocity and the $4\times 4$ ${\hat \gamma}$-matrices  
${\hat \gamma}_\mu=(i{\bf 1}\otimes{\sigma_3}, {\bf 1}\otimes{\sigma_2}, -{\sigma_3}\otimes\sigma_1)$ 
act in the space of the Dirac bi-spinors
$\psi=(\psi_K(A),\psi_K(B),\psi_{K^\prime}(A),\psi_{K^\prime}(B))$ 
composed of the values of the electron wave function on the $A$ and $B$ sublattices of the bipartite hexagonal lattice of graphene. 

The Dirac fermions are subject to random scalar (s) and vector (v) fields whose spatial correlations are controlled by the Gaussian averages
\be
\langle{a}_\mu({\bf q}){a}_\nu(-{\bf q})\rangle=
\delta_{\mu 0}\delta_{\nu 0}w_s({\bf q})+
\delta_{\mu i}\delta_{\nu j}w_v({\bf q})
\left( \delta_{ij}- {{q}_{i} 
{q}_j\over {\bf q}^2} \right)
\ee
The variances $w_{s,v}({\bf q})=4\pi^2{\Gamma_{s,v}/{\bf q}^2}$ are proportional to the areal densities of the Coulomb impurities ($\Gamma_s=g^2n_i$ where $g=e^2/\varepsilon_0v$ is the Coulomb interaction parameter) and topological defects ($\Gamma_v\sim n_d$), respectively. 

Strictly speaking, in the case of graphene the vector field 
$\vec {\bf a}$ becomes a $4\times 4$-matrix proportional to $({\bf 1}\otimes{\sigma_3})$ \cite{crespi}, thus endowing Eq.(2) with the structure of a $16\times 16$-matrix. It appears, however, that this technical complication does not affect any of the results presented below and, therefore, we will use Eq.(2) as is.

A preliminary insight into the problem can be gained by attempting to compute a fermion self-energy in the framework of the customary self-consistent Born approximation (SCBA)
\be
{\hat \Sigma}^R_{\alpha}(\epsilon,{\bf p})=\int {d{\bf q}\over (2\pi)^2}
{w_{\alpha}({\bf q})\over {\hat G}^R(\epsilon,{\bf p+q})^{-1}+{\hat \Sigma}^R_{\alpha}(\epsilon,{\bf p+q})}
\ee
In the case of RP, the singular behavior of 
$w_s({\bf q})$ at $q\to 0$ gets replaced by $w_s({\bf q})=\Gamma_s/(q+\kappa)^2$ 
due to the Debye screening. The corresponding Debye momentum $\kappa=4gp_F$ proportional 
to the Fermi surface radius $p_F=(\pi n_e)^{1/2}$, where $n_e$ is the 
density of excess electrons with respect to half-filling, renders the corresponding quasiparticle width finite
\be
\gamma_s=Im Tr{\hat \gamma}_0{\hat \Sigma}^R_s(\epsilon,\epsilon/v)\sim{v^2\Gamma_s\over\epsilon}min[{1\over g}, {1\over g^2}]
\ee
By contrast, a calculation of
the self-energy associated with RMF is impeded by the lack
of screening for the random vector potential which results in an infrared divergence 
\be
\gamma_v=Im Tr{\hat \gamma}_0{\hat \Sigma}^R_v(\epsilon,\epsilon/v)\sim {v\Gamma^{1/2}_v}\sqrt{\ln L}
\ee
where $L$ is the size of the system. 

This intrinsic divergence can not be avoided even if one proceeds beyond the SCBA, for it stems from the non-gauge invariant nature of the fermion Green function ${\hat G}^R(t,{\bf r})$ for ${\bf r}\neq {\bf 0}$. Taken at its face value, the divergent self-energy (5) is indicative of a strongly non-Lorentzian form of the RMF-averaged Green function \cite{altshuler}.

It is worth mentioning that, unlike in the previously discussed examples of the RMF problem involving some auxiliary fermions which are different from physical electrons \cite{hlr,RMF,franz}, in the present case there is no physical ground for replacing the original Green function (1) with a gauge-invariant amplitude such as, e.g., 
${\hat {\cal G}}^R(t,{\bf r})=<\psi(t,{\bf r})\exp(-i\int_{C}{\bf a}({\bf r}^\prime)d{\bf r}^\prime ){\psi}^\dagger(0,{\bf 0})>$.

Nonetheless, Eq.(3) can be used to evaluate the Drude transport rates in both, RP and RMF, cases. 
Inserting the factor $1-\cos\theta$ related to 
the transferred momentum $q=2p\sin\theta/2$ into the integrand in Eq.(3), and putting $\epsilon=\epsilon_F=vp_F$,
one obtains the first-order Born estimates
\be
\gamma^{tr}_{\alpha}=\int{d{\bf q}\over (2\pi)^2}\delta(\epsilon_F-v|{\bf p+q}|)w_\alpha({\bf q})\sin^2\theta
\ee
which both turn out to be finite
\be
\gamma^{tr}_{s}\sim {v^2\Gamma_s\over \epsilon_F}min[1, {1\over g^2}],
~~~ \gamma^{tr}_{v}\sim {v^2\Gamma_v\over \epsilon_F}
\ee
and inversely proportional to $n_e^{1/2}$, thereby giving rise to the 
Drude conductivity $\sigma\propto n_e$, in agreement with experiment \cite{exp} 
(note that the experimentally relevant value of the Coulomb coupling is $g\sim 1$) and in contrast to the situation in the conventional 
2DEG where $\gamma^{tr}_\alpha\sim\Gamma_\alpha/m$ contain the band mass $m$, thus yielding $\sigma\propto n^{1/2}_e$. 

To make a further progress, we employ a path-integral representation of the Dirac fermion Green function which was
devised in Ref.\cite{Stefanis} and has been previously applied to the problem of quasiparticle transport in the vortex line liquid phase of the cuprates \cite{ayash}
\bea
{G}^R(\epsilon,{\bf r}| {a}_\mu({\bf r}))=
\int^\infty_0 dt 
\int_{{\bf r}(0)=0}^{{\bf r}(t)={\bf r}}
D{\bf r}\,D{\bf p}
\cr
\exp[i{\hat S}_0(t)+i\int^t_0d\tau (a_0({\bf r})-
{d{\bf r}\over d\tau}({\bf A}+{\bf a}({\bf r})) ]
\eea
where $A_\mu$ represents external field (if any) and the free Dirac action reads
\be
{\hat S}_0(t)=
\int^t_0d\tau
[\epsilon{\hat \gamma}_0+{\bf p}({d{\bf r}\over d\tau}-{\hat {\bf \gamma}})]
\ee
In Eq.(8), the usual ordering of ${\hat \gamma}$-matrices with respect
to the proper time $\tau$ must be performed according to the order of their appearance in the series expansion of the exponent.
In contrast to various approximate (e.g., Bloch-Nordsieck) 
representations, the integration over the momentum 
${\bf p}(\tau)$ conjugate to the spatial coordinate ${\bf r}(\tau)$ allows one to account for the spinor structure of the fermion propagator (1) exactly. 

Averaging over the disorder variables introduces a product of the "Debye-Waller" attenuation factors into the integrand in Eq.(8) (here $u_\mu=(1, d{\bf r}/d\tau)$)
\bea
W_{\alpha}[{\bf r}(\tau)]=\exp[-{1\over 2}\int {d{\bf q}\over (2\pi)^2}
\int^t_0 d\tau\int^t_0 d\tau^\prime
\cr 
{\bf u}_\mu(\tau){\bf u}_\nu(\tau^\prime)\langle{a}_\mu({\bf q}){a}_\nu(-{\bf q})\rangle
e^{i{\bf q}({\bf r}(\tau)-{\bf r}(\tau^\prime))}]
\eea
Despite its generally gauge-non-invariant nature, the amplitude (8) does appear to be gauge-invariant for closed trajectories (${\bf r}=0$), thus allowing one to analyze such magneto-transport effects as dHvA (SdH) oscillations of the density of states (hence, magnetization, etc.) and conductivity in a weak uniform external magnetic field $B\ll \Gamma_{s,v}$.

Evaluating Eq.(8) on the semiclassical trajectories
which dominate the path integral and correspond to multiple repetitions of the Larmor orbit of radius $R_c=\epsilon/vB$, one obtains  
a field-dependent density of states 
\be
\nu(\epsilon |B)=\nu(\epsilon |0)\sum_{n=-\infty}^\infty
e^{2\pi inA(\epsilon)-n^2\delta S_1(\epsilon)}
\ee
where  $A(\epsilon)=\pi\epsilon^2/B$ is the area of the Larmor orbit and the attenuation factor contributes as
\be
\delta S_1(\epsilon)=-\sum_{\alpha=s,v}\ln W_\alpha=
\pi[{\Gamma_s\over B}+{\Gamma_v\epsilon^2\over v^2B^2}]
\ee
which allows one to identify the proper cyclotron rates
\be
\gamma^{cycl}_{s}\sim v\Gamma^{1/2}_s,~~~ 
\gamma^{cycl}_{v}\sim(\epsilon^2v^2\Gamma_v)^{1/4}
\ee
associated with the linear and quadratic (in powers of $1/B$)
Dingle plots, respectively. 

These results should be contrasted with their "non-relativistic"
counterparts ($\gamma^{cycl}_{s}\sim\Gamma_s/m,~~~ 
\gamma^{cycl}_{v}\sim(\epsilon\Gamma_v/m)^{1/2}
$). In combination with the transport rates (7), distinct energy (hence, electron density) dependences
of the rates (13) make it possible not only to distinguish between
the conventional 2DEG and graphene, but also
to discriminate between the RP and RMF mechanisms of elastic scattering (experimentally measured rates (13) would be naturally estimated for $\epsilon=\epsilon_F$).

With the use of the Poisson formula, Eq.(11) can be cast in the form
\be
\nu(\epsilon |B)\propto\sum_{n=0}^\infty\exp[-\pi{(\epsilon^2-\omega^2_n)^2\over v^2B(\gamma_s^{cycl})^2+(\gamma_v^{cycl})^4}]
\ee
showing an inhomogeneous broadening of the relativistic Landau levels
positioned at $\omega_n=\pm v(2Bn)^{1/2}$.

The path integral technique is also well suited for analyzing 
various gauge-invariant two-particle amplitudes. 
For one, the ballistic Drude conductivity manifesting the transport rates (7) can be found by computing the 
average $<{\hat {G}}^R(\epsilon, {\bf r}){\hat {G}}^A(\epsilon, -{\bf r})>$ in the semiclassical 
approximation. Expanding the corresponding path integral about a proper semiclassical trajectory (see below), 
one can systematically study ballistic corrections to the Drude conductivity \cite{dvk}. Due to the long-range nature
of the RMF disorder, these corrections do not appear to be logarithmic and, therefore, are not readily amenable to a simple resummation via
renormalization group (cf. with the case of short-range disorder \cite{aleiner}).
 
Another example of a relevant two-particle amplitude is given by the correlation function of the electron wavefunctions' amplitudes
related to the average \cite{gornyi} 
\bea
<{\hat {G}}^R(\epsilon, {\bf r}){\hat {G}}^R(\epsilon, -{\bf r})>=
\prod_{i=1,2}\int^\infty_0 dt_i 
\int_{{\bf r}_i(0)={\bf 0}}^{{\bf r}_i(t_i)=\pm{\bf r}}
D{\bf r}_iD{\bf p}_i
\cr
e^{i{\hat S}_0(t_1)}e^{i{\hat S}_0(t_2)} 
\prod_{\alpha=s,v}\prod_{i,j=1,2}W_\alpha({\bf r}_i(\tau_1)-{\bf r}_j(\tau_2))~~~
\eea
Observe that the product of the $W_v$-factors yields an exponent of the
Amperian area of a contour formed by the
trajectories ${\bf r}_1(\tau)$ and  $-{\bf r}_2(\tau)$.

In the ballistic regime, the integral (15) receives its main contribution 
from pairs of trajectories with single-valued projections onto the semiclassical
straight-path trajectory ${\bf r}^{(0)}(\tau)={{\bf r}}\tau/t$.
 
Proceeding by analogy with the earlier calculations in the case of the conventional 2DEG \cite{RMF,mirlin},
we separate out the coordinate variables onto the "center of mass" and relative ones 
(${\bf r}^{\pm}={\bf r}_1\pm {\bf r}_2$, ${\bf p}^{\pm}={\bf p}_1
\pm {\bf p}_2$), expand up to the second order in 
${\bf r}^{-}$ and ${\bf p}^{-}$, and 
integrate over all the variables except for the transverse deviation from the straight-path 
$x^{-}_\perp(\tau)$, thus arriving at the disorder-induced 
correction to the free Dirac action 
\be
\delta S_2(t)=\int^{t}_0 d\tau [\Gamma_s\epsilon
(x^{-}_\perp(\tau))^2+\Gamma_vv|x^{-}_\perp(\tau)| ]
\ee
Comparing typical values of the 
free fermion action and the correction (16), 
we find that the condition $S_0\gg\delta S$, under which the path integral (15) would be dominated by the trajectories close to ${\bf r}^{(0)}(\tau)$, is readily satisfied in the ballistic regime 
($\epsilon\gg v\Gamma_\alpha^{1/2}$).

As in the previous work \cite{RMF,ayash,gornyi}, it can be shown that the amplitude (15) is given by a
product of the free Dirac propagator with
$Im {\hat G}^R(\epsilon,{\bf r})\approx\epsilon({\bf 1}+{\hat {\bf \gamma}}{\hat {\bf r}})
J_0(\epsilon r)$ and
a Fourier transfrom (over the variable $\epsilon-p$)
of the Green function $g(x,x^\prime |\epsilon, p)|_{x=x^\prime=0}$ of the 
1D equation of motion in the direction perpendicular to the classical trajectory
\be
[v^2\partial^{2}_{x} + (\epsilon+iv\Gamma_v|x|+i\epsilon\Gamma_sx^2)^2-v^2p^2]g(x, x^\prime |\epsilon, p)=\delta(x-x^\prime)
\ee
The imaginary effective potential appearing in (17) restrains the transverse Dirac fermion's motion, unlike a 
real potential which allows for the Klein tunneling. 

In the pure RP or RMF case, the solution of Eq.(17) can be presented in the form
\bea
g_s(0,0 |\omega)=
{1\over (v^6\epsilon^2\Gamma_s)^{1/4}}f_s({\omega\over v\Gamma_s^{1/2}}),
\cr
g_v(0,0 |\omega)=
{1\over (v^5\epsilon\Gamma_v)^{1/3}}f_v({\omega\epsilon^{1/3}\over v^{4/3}\Gamma_v^{2/3}})
\eea
In the ballistic regime, the scaling functions $f_{s,v}(z)$ approach the parabolic cylinder and Airy functions, respectively.

Computing a real-space asymptotic behavior of the irreducible part
of the wavefunction amplitudes' correlator 
\bea
L^4<|\psi^2({\bf r})\psi^2({\bf 0})|>-1={<Im{\hat {G}}^R(\epsilon, {\bf r})
Im{\hat {G}}^R(\epsilon, -{\bf r})>\over (\pi\nu(\epsilon))^2}
\cr 
\sim ({\gamma^{FO}_\alpha\over \epsilon^2r})^{1/2}\cos(2\epsilon r)e^{-r\gamma^{FO}_\alpha}~~~
\eea 
one can identify the rates 
which control the decay of its Friedel-type oscillations 
\be
\gamma^{FO}_s\sim {v\Gamma^{1/2}_s},~~~
\gamma^{FO}_{v}\sim v^{4/3}{\Gamma^{2/3}_v\over \epsilon^{1/3}}  
\ee
and contrast them against their non-relativistic counterparts
($\gamma^{FO}_s\sim{\gamma^{tr}_s}$ and
$\gamma^{FO}_{v}\sim \Gamma^{2/3}_v\epsilon^{1/3}/m^{2/3}$).  

These rates (equivalently, length scales) 
can also be manifested by other types of Friedel oscillations,
such as that of the electron density profile induced by an isolated impurity or the RKKY interaction between a pair of magnetic ions.
Namely, at distances $r\sim v/\gamma^{FO}_\alpha$ the previously found behavior $\delta\rho(r)\propto\cos(2\epsilon r)/r^3$
\cite{stauber} changes to
\be
\delta\rho(r)\propto({\gamma^{FO}_\alpha\over r^{5}})^{1/2}\cos(2\epsilon r)e^{-r\gamma^{FO}_\alpha}
\ee
In the complementary low energy limit ($\epsilon\lesssim v\Gamma^{1/2}_\alpha$), the above eikonal-type approach ceases to be applicable.
Nonetheless, one can still gain some insight into the localization properties of the system in question by focusing on zero-energy states (if any).

In the (apparently, more challenging) case of a pure RMF, these states can be explicitly constructed in the form
\be
\psi_{\pm}({\bf r})\propto ({\bf 1}\pm {\hat \gamma}_0)
\pmatrix{e^{\phi({\bf r})}
\cr
e^{-\phi({\bf r})}} 
\ee
for an arbitrary configuration of the random vector potential 
parameterized as ${a}_i({\bf r})=\epsilon_{ij}\nabla_j\phi({\bf r})$. 

In a finite system, a degree of the wave functions' localization (or a lack thereof)  
can be inferred from the inverse participation ratios 
\be
{\cal P}_n=
<{\int |\psi({\bf r})|^{2n}d{\bf r}\over {L^2(\int |\psi({\bf r}^\prime)|^2d{\bf r}^\prime)^n}}>
\sim{\Gamma_v^{n-1}\over L^{2}}
\ee
where
the Gaussian averaging over the disorder 
field $\phi({\bf r})$ was performed with the weight  
$P[\phi({\bf r})]\propto\exp(-\int d{\bf r}(\nabla^2\phi)^2/2\Gamma_v)$. 

The expression (23) is in a stark contrast with that in the short-range
case where the zero-energy wave functions demonstrate a "pre-localized" behavior and the participation ratios exhibit a multifractal spectrum of anomalous dimensions ${\cal P}_n\propto L^{-an+bn^2}$ \cite{Ludwig}. 

Eq.(23) is suggestive of a strong localization of the 
zero-energy states, the apparent localization length being of order $\sim{\Gamma_v}^{-1/2}$.
Moreover, it is conceivable that all the states
up to the energy of order $\sim v\Gamma^{1/2}_v$
remain localized. It should be noted, however, that 
by modeling RMF as a random variable with the variance (2) one misses out on the possibility of percolating "snake" trajectories, akin to those found in the case of non-relativistic fermons in a smoothly varying RMF with the correlation length greater 
than $v/\epsilon$ \cite{polyakov}. 
 
As far as two-particle processes are concerned,
a diffusive behavior of the corresponding amplitudes for $|\epsilon_1\pm\epsilon_2|\lesssim\gamma^{tr}_\alpha$) can be studied in a more familiar framework of the non-linear $\sigma$-model (NL$\sigma$M). 
In the metallic regime ($\epsilon_F\gg\gamma^{tr}_\alpha$), 
a derivation of the corresponding (supersymmetric) NL$\sigma$M 
would closely follow the solution of the RMF problem 
for non-relativistic spinful fermions with a gyromagnetic ratio equal two \cite{kogan}. This model features the same
unitary symmetry as that with only an orbital coupling to the
long-range RMF \cite{sigma}.

In the opposite ("undoped") limit,
$\epsilon_F\lesssim\gamma^{tr}_\alpha$, a systematic derivation of the 
NL$\sigma$M is hindered by the lack of a suitable expansion parameter,
for the bare conductivity takes values of order unity (in units of $\sim e^2/h$). For methodological purposes, however, one can still introduce a large number of valleys $N$, similar to the analysis of dirty $d$-wave superconductors with $\epsilon_F=0$, thereby allowing for a formal $1/N$-expansion.

This caveat notwithstanding, it was recently argued 
that in the case of short-range disorder the NL$\sigma$M obtained by non-abelian bosonization
could hold all the way down to the lowest energies \cite{altland}. 
It would be interesting to see if this conjecture can also be extended to a long-range RMF, in which case the corresponding localization scenario would fall into the so-called "$C I$" universality class
\cite{zirnbauer}. 
Such a behavior develops in the presence of a non-vanishing inter-valley scattering (due to, e.g., a screened RP)
in addition to the small-angle (intra-valley) one described by Eq.(2).
It is, however, less likely that even in the absence of inter-valley scattering the system would show any scale-invariant behavior (universality class $A$ \cite{zirnbauer}), as it does in the short-range case \cite{Ludwig}.

In light of the above, the argument of Ref.\cite{morpurgo} that RMF scattering suppresses any quantum coherence between pairs of time-reversed trajectories (which can result in either localizing or antilocalizing behavior, depending
on the relative strength of the intra- vs inter-valley scattering \cite{ando}) would only apply in the case of a smoothly varying RMF.
As the aforementioned NL$\sigma$M analyses suggest, in the case of RMF described by Eq.(2) the onset of localization gets merely postponed until greater length scales ($e^{\pi\sigma^2}v/\gamma^{tr}_v$ instead of $e^{\pi\sigma}v/\gamma^{tr}_v$, provided that the bare conductivity $\sigma\gg 1$). 

In summary, we study the behavior  
of two-dimensional Dirac fermions subject to long-range-correlated random scalar and vector potentials. 
In the ballistic regime of large quasiparticle energies, we 
obtain estimates of the scattering rates 
manifested by the Drude conductivity, dHvA/SdH and Friedel oscillations. The distinct energy (density) dependencies of such rates provide a means of ascertaining the dominant mechanism of elastic scattering in graphene. 
In the complementary low-energy regime, we 
find a signature of strong localization in a long-range RMF and discuss pertinent localization scenarios.

This research was supported by NSF under Grant DMR-0349881.

\wt
\end{document}